\documentclass[3p,times]{elsarticle}

\usepackage{ecrc}

\volume{00}

\firstpage{1}

\journalname{Nuclear Physics A}

\runauth{J. Schukraft}

\jid{nupha}

\usepackage{amssymb}

\usepackage[figuresright]{rotating}

\newcommand{\nGc}{GeV/$c$}
\newcommand{\nY}{$\Upsilon$}
\newcommand{\pt}{$\rm p_T$ }
\newcommand{\raa}{$\rm R_{AA}$ }
\newcommand{\npi}{$\pi$\ }

\begin{document}

\begin{frontmatter}



\dochead{}

\title{Hard Probes 2012: Experimental Summary}

\author{J. Schukraft\fnref{label2}}
\fntext[label2]{Email: schukraft@cern.ch}

\address{CERN, Div. PH, CH-1211 Geneva 23}

\begin{abstract}
The \emph{$5^{th}$ international Conference on Hard and Electromagnetic Probes in High-Energy Nuclear Collisions} was held in May 2012 in Cagliari, Italy. This contribution summarises some of the experimental highlights presented at the meeting, concentrating on new results from LHC and RHIC on parton energy loss ('jet-quenching') and heavy quark meson production ('quarkonia suppression').

\end{abstract}


\end{frontmatter}


\section{Introduction}
The Hard Probes 2012 conference was the first in this series since the start of the ion program at the CERN LHC about 1.5 years ago. It saw a wealth of new and high precision data from LHC, including first results from the 2011 high luminosity ion run and pp comparison data from different centre-of-mass energies, as well as plenty of fresh and significant results from RHIC, in particular on heavy flavours. This experimental summary will mention briefly a few of the highlights and the reader is referred to the individual experimental contributions in these proceedings for details and all original plots and results (all contributions to these proceedings are referred to in the text by listing speaker/experiment). Recent reviews of ultra-relativistic heavy ion physics can be found in~\cite{ Muller:2006ee,Jacak:2012dx, Muller:2012zq}.

\section{Proton-proton and proton-nucleus comparison data}
Many heavy ion results rely on a direct comparison to pp and pA data; high precision 'comparison data' is therefore crucial, if not always flashy and glamorous. Besides its prime purpose of providing the baseline against which to quantify the modifications from either cold (nuclear) or hot (QGP, hadron gas) matter, comparison data is needed to tune event generators, help in establishing the validity range and accuracy of pQCD calculations, and last but not least, offers an important cross- and consistency check for detector performance and analysis techniques. The good news at this conference was therefore that plenty of pp comparison data was presented, including open charm cross sections from LHC (C. Geuna/ALICE) and RHIC (D. Tlusty/STAR); hidden charm (J/$\psi$, $\chi_c$ C. Geuna /ALICE, C. Lourenco/CMS, M. Calderon de la Barca Sanchez/STAR) and hidden beauty ($\Upsilon$ family, C. Lourenco/CMS, A. Kesich/STAR, S. Whitaker/PHENIX); jet cross sections and jet fragmentation (R. Ma/ALICE, D. Perepelitsa/PHENIX). The bad news was that the LHC pp data at the correct comparison energy of 2.76 TeV was taken in only a couple of days during 2010 at rather low luminosity and therefore has mostly much larger statistical errors than the corresponding nuclear data! This is, at the very least, unfortunate, and a few more days can hopefully be found in the LHC schedule during the ongoingupsilon
 pp run to remedy the situation.  

Proton-nucleus data is of equal importance for comparison to the nuclear data and interesting in its own right to study shadowing and saturation physics, e.g. in the theoretical framework of the Colour Glass Condensate model. The first long pA run at LHC will only happen later in early 2013,  but new d-Au data are available from RHIC and show a surprisingly large suppression of jets up to 30 GeV relative to the scaled pp expectations (D. Perepelitsa/PHENIX); naively one would have expected shadowing effects to be much smaller at the corresponding large values of $Q^2$.
The Upsilon at forward rapidity looks suppressed as much if not more than the J/$\psi$ (A. Frawley/PHENIX). Also this is contrary to expectation, based on fixed target pA results, that cold nuclear matter absorption should be smaller for the $\Upsilon$ than for the J/$\psi$. While the statistical significance of this result is not very high at this moment, it serves as a useful reminder that a measurement is always preferable to even a well-motivated expectation.  

\section{Energy loss}
The energy lost by energetic partons which traverse the hot and dense matter ('jet quenching') measures the 'stopping power' of the QGP, quantified by the transport coefficient $\hat{q}$, as well as its characteristic dependence on parton type (quark vs. gluon), parton mass $m$ (light vs. heavy quark), path length inside the medium $L$ , and parton energy $E$. Besides the obvious question of how much energy is lost, usually addressed by measuring the energy balance between di-jets or gamma-jet pairs, one has also to verify the scaling with $L, m,$ and $ E$, for which different models make different predictions. Finally, in order to learn about the dynamics of the energy loss process (e.g. few hard gluon versus multiple soft gluon radiation, contributions of collisional and radiative energy loss), one has to measure the energy and angular distribution of the 'lost energy', which reappears typically in the form of very soft particles ($<$ few \nGc) and at large angles to the initial parton direction. A number of new results, and addressing all of the above questions, have been shown at this conference in a variety of different observables, from inclusive single particle spectra (unidentified/identified/heavy flavour), two and three particle correlations on both near and away side, inclusive single jets ($R_{AA}$, longitudinal and transverse fragmentation functions), and finally jet correlations (di-jet, $\gamma(Z)$-jet). Only by looking at many observables one can tie together and characterise the different facets of jet-quenching (P. Jacobs/ALICE).

Suppression of charged hadrons, the signal where jet quenching was discovered at RHIC, is still a very useful and sensitive observable even in the era of fully reconstructed jets, and severely constrains the various models of jet quenching at LHC (M. Floris/ALICE). The suppression is found to be independent of particle species above some 8-10 \nGc~(P. Christiansen/ALICE), and the 2011 high luminosity run has significantly decreased the statistical errors at high pt (K. Krajczar/CMS), where the suppression remains at about a factor of two up to at least 100 \nGc.

Suppression of calorimetric (A. Angerami/ATLAS) or charged (M. Verweij/ALICE) jets is also strong out to the highest measured jet energy of about 200 GeV and of the same order ($R_{AA} \approx 0.5)$ as the suppression of single high \pt particles. This implies that very little or none of the lost energy is collected within the typically used jet-cones ($R < 0.3-0.4$, \pt $> 3-5$ \nGc). New high statistics data on di-jets (C. Roland/CMS) confirm that the back-to-back correlation does not change with increasing energy loss from its value measured in pp, implying that the medium induced transverse momentum kick is (much) smaller than the natural broadening from initial and final state vacuum radiation. The average energy imbalance between di-jets is about 10\% higher in central Pb-Pb than in pp, independent of jet energy out to more than 300 GeV. Together with the very first results on energy balance in the particularly clean $\gamma$-jet channel (Y. Lai/CMS), these data can be used to extract, via model comparisons, both the amount of energy loss (or $\hat{q}$) as well as its energy dependence. Also the first $\gamma$-hadron correlations measured at RHIC (J. Frantz/PHENIX) show the expected medium effect on fragmentation functions.

Electroweak probes are excellent 'calibration candles' for jet-quenching measurements, as both production and propagation should not be effected by the medium. And indeed, as was amply demonstrated at this conference, it isn't (P. Steinberg/ATLAS, Z. Citron/ATLAS, A. Milov/ATLAS, B. De la Cruz/CMS, Y.J. Lee/CMS, L. Benhabib/CMS). The nuclear modification factor \raa is one, within current systematic and statistical errors of order 10-20\%, for all measured centralities, transverse momenta, and probes ($\gamma$, W, Z). These results confirm that the collision geometry and assorted derived parameters like impact parameter, number of participants and collisions, are under good experimental control (shadowing effects at the relevant $Q^2$ are expected to be below the 5-10\% level).

The fate of heavy (c,b) quarks in the QGP is of interest in several respects. They are tracers created essentially only during the initial collision in hard scattering processes and their energy loss should be smaller than the one of gluons (different colour charge) or of light quarks (mass dependent reduction of energy loss or 'deadcone' effect). In a very strongly interacting plasma, they may eventually become thermalized and take part in the hydrodynamic collective flow of the bulk matter. 

Heavy flavour results from LHC and RHIC include fully reconstructed D-mesons (Z. Conesa del Valle/ALICE, M. Calderon de la Barca Sanchez/STAR), heavy flavour muons (Y. Chen/ATLAS, D. Stocco/ALICE, K. Read/PHENIX) and electrons (M. Kweon/ALICE, O. Hajkova/STAR, S. Tarafdar/PHENIX), as well as B-mesons identified via non-prompt J/$\psi$ decays (T. Dahms/CMS). Overall indications are roughly consistent with the expected hierarchy of suppression, \raa(\npi) $\le$ \raa(D) $<$ \raa(B); however statistical and systematic errors have to be improved and potential shadowing corrections need to be quantified with pA data before definite conclusions can be drawn. The LHC data  however do confirm with fully reconstructed charm mesons (S. Masciocchi/ALICE) what has been seen previously with heavy flavour ('non-photonic') leptons at RHIC, namely that the suppression of charm is very similar to that of other hadrons above 5-8 \nGc~and larger only, if at all,  for non-relativistic momenta ($p_T/m$ of order one).

Indications, again at the edge of significance, have been shown for charm elliptic flow from both LHC and RHIC (G. Ortona/ALICE, O. Hajkova/STAR, S. Tarafdar/PHENIX) for momenta above 2 \nGc, of a magnitude comparable to the one of charged particles. This could be a sign of charm quark thermalisation at the lower momenta and/or path length dependent energy loss at the higher momenta; smaller error bars and an extension towards higher momenta at RHIC and lower momenta at LHC would be very helpful for a quantitative interpretation.

In order to investigate the energy lost by the jet-quenching process, experimentalists have to dig into the soft part of the hadron spectrum, typically by looking at correlations between a trigger particle (as a proxy for the jet) and soft associated particles of a few GeV. Trigger bias, which in this method enriches the sample with unmodified jets from the surface, as well as correlations introduced by hydrodynamic flow, have to be taken into account in analysis and interpretation. Data on two particle correlations from LHC (J.F Grosse Oetringhaus/ALICE) and RHIC, (A. Ohlsen/STAR, J. Franz/PHENIX) cover the interesting momentum range from around 1 \nGc~to above 10 \nGc, i.e. the region where one expects the lost energy to reappear  as well as any medium response to the large local energy deposited by jet quenching. Indeed, tell-tale signatures of increased soft particle correlations are seen in the energy balance ($D_{AA}$, A. Ohlsen/STAR) and the azimuthal correlation (T. Todoroki/PHENIX) at the away side at RHIC. At LHC, the near-side shows a curious modification (A. Morsch/ALICE), which affects the width of the rapidity correlation but not (or much less) the width of the azimuthal correlation between trigger and associated particles. Whether this is a sign of medium modified jet fragments (radiated jet energy) or jet modified medium fragments (response of the medium to local energy deposit) remains to be seen; interestingly enough the AMPT Monte Carlo shows quantitative agreement with the data without, so far at least, revealing its underlying cause.

Further hints about the nature of jet quenching come from correlations with identified particles, where it seems that the anomalous large proton to pion ratio typical of heavy ion collisions is not present in jet fragments on the near side (M. Veldhoen/ALICE) and is significantly reduced on the away side (A. Davila/STAR). While the former could be caused, at least in part, by the previously mentioned surface bias, the latter clearly is not. These results may therefore indicate that jet fragments are less subject to bulk effects like radial flow and coalescence, both of which are thought to be the cause of the baryon enhancement in bulk matter. 

\section{Quarkonia suppression}
The melting of hidden heavy quark mesons (the J/$\psi$ and \nY~families) is arguably the most prominent and sought after signal of deconfinement in the QGP. Notwithstanding its 1986 declaration as an 'unambiguous signature of QGP formation', and the first observation of J/$\psi$ suppression the same year by the NA38 collaboration at the CERN SPS, J/$\psi$ production in nuclear collisions has surprised and confounded the heavy ion community for more than 25 years: from the first suppression measurement with light ions, later attributed to hadronic final state absorption in cold nuclear matter, the 'anomalous' J/$\psi$ suppression seen later in central Pb-Pb collisions, the indecisive results with lighter Indium ions, to the initially very surprising observation that J/$\psi$ suppression as function of centrality at the SPS is essentially identical to the one measured at more than ten times higher energy at RHIC. For the latter observation, two broad classes of explanations were put forward. In the sequential suppression scenario, SPS and LHC results are 'equal by design', because the temperature at both machines remains below the critical temperature needed to melt the directly produced J/$\psi$, but is (in both machines) above the melting temperature of $\psi$' and $\chi_c$, which together contribute somewhat less than 1/2 of the total J/$\psi$ yield. In the recombination picture, the suppression caused by melting (deconfinement) of all members of the  J/$\psi$ family increases with temperature from SPS to RHIC, but is approximately cancelled by secondary J/$\psi$ production via coalescence between independently created charm quarks; in this scenario SPS and RHIC are 'equal by accident'. A decisive answer was (yet again) expected by going to even higher energy, the LHC, because melting and recombination in general should depend differently on $\sqrt{s}$. The sequential suppression picture predicted an equal or stronger suppression at LHC, the recombination picture a weaker one (or even an enhancement).

And indeed, the LHC lived up to expectation. High statistics results shown for the first time at this conference (C. Suire/ALICE) establish that the J/$\psi$ is overall \emph{less} suppressed at LHC than at RHIC, but that, unlike at RHIC, the suppression is \emph{stronger} at high \pt than at low \pt. In addition, indications are that the J/$\psi$ shows elliptic flow at LHC (L. Massacrier/ALICE), but not at RHIC (M. Calderon de la Barca Sanchez /STAR). This pattern is qualitatively and, within uncertainties, quantitatively consistent with the coalescence prediction that the abundantly produced individual charm quarks at LHC thermalize and recombine, and preferentially so at low relative momenta. While at first sight regeneration via charm quark coalescence may seem to be yet another 'dirty' effect which obscures the deconfinement signal, it actually is a consequence of deconfinement itself and therefore a very relevant observation! Unlike the light quarks, charm quarks are produced at the very beginning of the collisions in independent (and distant) hard interactions; they therefore have to roam freely and over large distances to occasionally find a coalescence partner to bind to; in other words, they are deconfined during the system evolution. 

The story seems to be more complicated for the $\psi$', which was supposed to have melted away completely already at RHIC. In a totally unexpected and bizarre twist, the $\psi$' was found at LHC to be \emph{enhanced} with respect to the J/$\psi$ at low \pt(T. Dahms/CMS)! If this 'anomalous $\psi$' un-suppression', which as of now has limited significance due to the lack of 2.76 TeV pp comparison data, stands the test of time, it will have totally unpredictable, but definitely profound consequences for the quarkonia suppression!

Recombination is thought to be less of an issue for beauty quarks, and therefore \nY~suppression should reflect the resonance melting more directly. The pattern observed at LHC (C. Mironov/CMS) is the following: the tightly bound \nY(1S) is suppressed by slightly more than a factor of two, the \nY(2S) by about a factor of ten, and the most loosely bound state, the \nY(3S), is no longer visible within statistics. This is qualitatively consistent with a 'sequential' melting scenario, where most of the higher mass resonances have partially or completely disappeared whereas the directly produced \nY(1S) may actually be not suppressed at all (as about 50\% of the observed \nY(1S) state results from feed down from $\chi_b$ and other higher resonance states). 

Crucial support for a sequential melting scenario would come from observing a change of the suppression pattern with temperature (i.e. beam energy). First results on the sum of the three S-wave \nY~states have been shown from RHIC (A. Kesich/STAR, S. Whitaker/PHENIX); the suppression is comparable to the one at LHC, maybe a bit smaller (which would be in line with a melting scenario), but errors are too large at the moment and this critical comparison has to await more statistics from RHIC.

Finally, and for the first time since many years, the quarkonia results seem qualitatively consistent with the presence of a deconfining medium, which implies both melting as well as regeneration of some of the heavy quark bound states. However, even leaving aside for the time being the $\psi$', a quantitative understanding needs better statistics (in particular for the \nY), correcting for possible cold nuclear matter effects (absorption/shadowing) at LHC with data from the upcoming pPb run, and quantitatively taking into account (if possible, even measuring) the contributions of feed down from the various higher lying quarkonia resonance states.

\section{Flow and electromagnetic radiation}
While not a traditional focus at the Hard Probes conferences, a number of significant new results were reported on flow observables, mostly from the intermediate 'semi-hard' momentum range.

 Azimuthal anisotropies (v$_2$) become independent of particle species (P. Christiansen/ALICE), but only above some 10 \nGc; presumably at these high momenta they are no longer caused by hydrodynamic flow. The anisotropy extends out to an amazing 50 \nGc~(C. Maguire/CMS), with a magnitude roughly in line with the predictions from path length dependent jet-quenching calculations. Elliptic flow fluctuations have been measured for the first time as a function of transverse momentum (J.F Grosse Oetringhaus/ALICE), and are found to be amazingly independent of it, from the lowest momenta where hydro flow dominates ($<$ 3 \nGc) out to the region presumably governed by the completely different mechanism of jet-quenching ($>$ 6 \nGc). A first measurement of the rapidity-even directed flow (v$_1$) has been reported (S. Mohapatra/ATLAS), taking into account momentum conservation terms, as well as a remarkable variety of higher order flow plane correlations (J. Jia/ATLAS), some of which seem caused by subtle correlations in the initial geometry, others by non-linear evolution in the hydro expansion. 

A significant excess is seen in the low mass electron pair continuum ($m_{ee} \approx$ 200 - 800 MeV/$c^2$) at RHIC (L. Ruan/STAR, A. Drees/PHENIX). This most important signal, of relevance for both thermal radiation from the QGP as well as chiral symmetry restoration at the phase boundary, is also one of the most difficult signals to observe in heavy ion collisions, because of the very small signal-to-background ratio. While both experiments report a significant excess above standard resonance contributions, the two results are unfortunately also known to be inconsistent with each other, with the larger value (from PHENIX) well above any current  estimate of thermal radiation and/or rho-meson broadening. A resolution of this discrepancy is eagerly awaited, as are first results from LHC, which however will be even more difficult to extract than at RHIC. 
 
\section{Summary}
Less than 18 months into heavy ion operation at LHC, and with a steady stream of new and very competitive results from RHIC, the 'Hard Probes 2012' conference has seen impressive progress for the field. On the topic of parton energy loss/jet-quenching, some of the expected scaling laws -- mass, energy, path length, colour charge -- are being put to the test, with passing grades so far (some only barely); first hints are found of the lost energy in modified fragmentation functions and two particle correlations; and the 'holy grail' signal of clean $\gamma(Z)$-jet pairs is starting to appear, begging for more statistics and inviting direct comparisons with theory.  Light is visible at the end of the quarkonia tunnel with the  tentative picture taking hold of both sequential quarkonia melting as well as quarkonia recombination, both direct consequences of deconfinement, but also a potential new obstacle in the form of a ghostly resurrected $\psi'$.  In the bulk matter below 2-3 \nGc, hydro reigns supreme not only to leading (elliptic and radial flow) and next-to-leading (higher harmonic flow and fluctuations) order, but even to NNLO, e.g. explaining some of the non-trivial correlations between different order reaction planes as consequences of non-linear hydrodynamic evolution. The intermediate momentum region however, joining the soft and the hard regimes between a few and some ten \nGc, remains the least explored and the least understood, and it is to be hoped that in a future hard probes conference also this region will be coming more into focus from both experiment and theory.










\end{document}